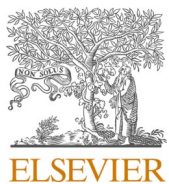
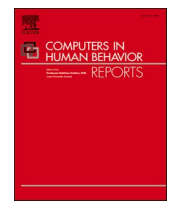

# How immersive virtual reality methods may meet the criteria of the National Academy of Neuropsychology and American Academy of Clinical Neuropsychology: A software review of the Virtual Reality Everyday Assessment Lab (VR-EAL)

Panagiotis Kourtesis [a,b,c,d,*], Sarah E. MacPherson [e,f]

[a] *National Research Institute of Computer Science and Automation, INRIA, Rennes, France*
[b] *Univ Rennes, Rennes, France*
[c] *Research Institute of Computer Science and Random Systems, IRISA, Rennes, France*
[d] *French National Centre for Scientific Research, CNRS, Rennes, France*
[e] *Human Cognitive Neuroscience, Department of Psychology, University of Edinburgh, Edinburgh, UK*
[f] *Department of Psychology, University of Edinburgh, Edinburgh, UK*

ARTICLE INFO

ABSTRACT

Clinical tools involving immersive virtual reality (VR) may bring several advantages to cognitive neuroscience and neuropsychology. However, there are some technical and methodological pitfalls. The American Academy of Clinical Neuropsychology (AACN) and the National Academy of Neuropsychology (NAN) raised 8 key issues pertaining to Computerized Neuropsychological Assessment Devices. These issues pertain to: (1) the safety and effectivity; (2) the identity of the end-user; (3) the technical hardware and software features; (4) privacy and data security; (5) the psychometric properties; (6) examinee issues; (7) the use of reporting services; and (8) the reliability of the responses and results. The VR Everyday Assessment Lab (VR-EAL) is the first immersive VR neuropsychological battery with enhanced ecological validity for the assessment of everyday cognitive functions by offering a pleasant testing experience without inducing cybersickness. The VR-EAL meets the criteria of the NAN and AACN, addresses the methodological pitfalls, and brings advantages for neuropsychological testing. However, there are still shortcomings of the VR-EAL, which should be addressed. Future iterations should strive to improve the embodiment illusion in VR-EAL and the creation of an open access VR software library should be attempted. The discussed studies demonstrate the utility of VR methods in cognitive neuroscience and neuropsychology.

## 1. Introduction

A series of studies from our laboratory have adopted a multidisciplinary approach (i.e., computer science and psychology) to explore the potency of immersive virtual reality (VR) as a research and clinical tool in cognitive neuroscience and neuropsychology. The studies have also addressed the issue of ecological validity in neuropsychological testing, especially regarding the assessment of cognitive functions which are central to everyday functioning. Finally, we have examined the technical and methodological pitfalls associated with the implementation of immersive VR in cognitive neuroscience and neuropsychology.

Firstly, a technological systematic literature review of the reasons for adverse VR induced symptoms and effects (i.e., cybersickness) was conducted (Kourtesis, Collina, Doumas, & MacPherson, 2019a). The review provided suggestions and technological knowledge for the implementation of VR head-mounted displays (HMD) in cognitive neuroscience (Kourtesis et al., 2019a). A meta-analysis of 44 neuroscientific and neuropsychological studies involving VR HMD systems was also performed. Another aim was to devise a brief screening tool to quantitatively appraise and report both the quality of software features and cybersickness intensity, as such a tool did not exist. The Virtual Reality Neuroscience Questionnaire (VRNQ) was developed and validated to assess the quality of VR software in terms of user experience, game mechanics, in-game assistance, and cybersickness (Kourtesis,

Collina, Doumas, & MacPherson, 2019b). The same study provided suggestions pertaining to the maximum duration of VR sessions (Kourtesis et al., 2019b).

Guidelines were also proposed that described the development of the Virtual Reality Everyday Assessment Lab (VR-EAL), the first immersive VR neuropsychological battery, programmed using Unity game development software (Kourtesis, Korre, Collina, Doumas, & MacPherson, 2020b). Furthermore, the convergent, construct, and ecological validity of VR-EAL as an assessment of prospective memory, episodic memory, visual attention, visuospatial attention, auditory attention, and executive functions were examined (Kourtesis, Collina, Doumas, & MacPherson, 2020a). Finally, using VR-EAL, prospective memory in everyday life was examined by comparing performance on diverse prospective memory tasks (i.e., focal and non-focal event-based, and time-based tasks; Kourtesis, Collina, Doumas, & MacPherson, 2021) and identifying the cognitive functions which predict everyday prospective memory functioning (Kourtesis & MacPherson, 2021).

The findings of these aforementioned studies have already been published as individual studies. However, the results of these studies will be discussed here using an all-inclusive approach in an attempt to examine whether the VR-EAL meets the criteria of the National Academy of Neuropsychology (NAN) and American Academy of Clinical Neuropsychology (AACN) for Computerized Neuropsychological Assessment Devices (CNADs). The VR-EAL was designed and developed to meet the NAN and AACN criteria. The current software review will examine whether the VR-EAL indeed meets these criteria and offers a discussion of how other immersive VR CNADs may also meet them.

## 2. Summary of the VR-EAL and relevant studies

VR-EAL assesses everyday cognitive functions such as PM, episodic memory (i.e., immediate and delayed recognition), executive functioning (i.e., planning, multitasking) and selective visual, visuospatial and auditory (bi-aural) attention within a realistic immersive VR scenario lasting around 70 min. The VR-EAL offers both tutorials and a continuous storyline in an alternating fashion. See Table 1 and Figs. 1–4 for a summary of the VR-EAL scenario and tasks. A brief video recording of the VR-EAL may also be accessed at this hyperlink: https://www.youtube.com/watch?v=IHEIvS37Xy8&t.

VR-EAL can be run on any VR HMD which is compatible with SteamVR, such as the HTC Vive series (e.g., Pro, Pro Eye, and Cosmos), Oculus Rift series (e.g., Rift and Rift S), Pimax series (e.g., 4K, 5K, and 5K Plus), Varjo series (e.g., VR-1 and VR-2), Samsung Odyssey series (e.g., Odyssey and Odyssey +), and Valve Index. Other criteria that should be met for efficient implementation of VR-EAL include the size of the VR area, which should be 5 $m^2$ to provide an adequate space for immersion and naturalistic interaction within virtual environments (Borrego, Latorre, Alcañiz, & Llorens, 2018). The spatialized (bi-aural) audio should be facilitated by a pair of headphones and the HMD should be connected to a laptop with the following minimum characteristics: Intel i5-4590/AMD Ryzen 5 1500X or greater, NVIDIA GTX 1060/AMD Radeon RX 480 or greater, NVIDIA GTX 970/AMD Radeon R9 290 or greater, 8 GB+ RAM, and high definition audio.

The development and compatibility of VR-EAL was based on a systematic literature review (Kourtesis et al., 2019a) and a study on the acceptability of VR technologies (Kourtesis et al., 2019 ab). The suitability of VR-EAL was thoroughly examined during the development phase (Kourtesis et al., 2020b) and its validity and advantages were evaluated against an extensive paper-and-pencil neuropsychological battery (Kourtesis, Collina, Doumas, & MacPherson, 2020). The contribution of VR-EAL in the understanding of everyday cognitive functions was also examined (Kourtesis, Collina, Doumas, & MacPherson, 2021; Kourtesis & MacPherson, 2021). Table 2 provides a summary of the aims and the findings for each study included in this series of studies. As Table 2 illustrates, the implementation of immersive VR in cognitive neuroscience and neuropsychology may be efficient and advantageous. Specifically, it is feasible to avoid or substantially alleviate adverse cybersickness and provide a neuropsychological assessment like VR-EAL with enhanced ecological validity and a shorter administration time. Also, the VR-EAL was rated as a highly pleasant testing experience and able to contribute to the understanding of everyday cognition.

**Table 1**
VR-EAL scenario.

| Order | Type | Description |
|---|---|---|
| Scene 1 | *Tutorial* | Basic interactions and navigation |
| Scene 2 | *Tutorial* | Interactive boards (recognition and planning) |
| Scene 3 | *Storyline* | List of prospective memory tasks, shopping list (immediate recognition), and itinerary (planning) |
| Scene 4 | *Tutorial* | List of mechanics for the prospective memory tasks, prompts, and notes |
| Scene 5 | *Tutorial* | Cooking |
| Scene 6 | *Storyline* | Prepare breakfast (multi-tasking) and take medication (prospective memory, event-based, short delay) |
| Scene 7 | *Tutorial* | Tutorial: collect items |
| Scene 8 | *Storyline* | Collect items from the living-room (selective visuospatial attention) and take a chocolate pie out of the oven (prospective memory, event-based, short delay) |
| Scene 9 | *Tutorial* | Interaction with 3D non-player characters |
| Scene 10 | *Storyline* | Call Rose (prospective memory task, time-based, short delay) |
| Scene 11 | *Tutorial* | Gaze interaction |
| Scene 12 | *Storyline* | Detect posters on both sides of the road (selective visual attention) |
| Scene 13 | *Tutorial* | Shopping, how to collect the items from the supermarket |
| Scene 14 | *Storyline* | Collect the shopping list items from the supermarket (delayed recognition) |
| Scene 15 | *Storyline* | Go to the bakery to collect the carrot cake (prospective memory task, time-based, medium delay) |
| Scene 16 | *Storyline* | False prompt before going to the library (prospective memory task, event-based, medium delay) |
| Scene 17 | *Storyline* | Return the red book to the library (prospective memory task, event-based, medium delay) |
| Scene 18 | *Tutorial* | Auditory interaction |
| Scene 19 | *Storyline* | Detect sounds from both sides of the road (selective auditory attention) |
| Scene 20 | *Storyline* | False prompt before going back home (prospective memory task, time-based, long delay) |
| Scene 21 | *Storyline* | When you return home, give the extra pair of keys to Alex (prospective memory task, event-based, long delay) |
| Scene 22 | *Storyline* | Put away the shopping items and take the medication (prospective memory task, time-based, long delay) |

## 3. Meeting the criteria of the National Academy of Neuropsychology and the American Academy of Clinical Neuropsychology

Bauer and collaborators (2012) published the official joint position of the American Academy of Clinical Neuropsychology (AACN) and the National Academy of Neuropsychology (NAN) which discusses 8 key issues regarding the development, dissemination, and implementation of Computerized Neuropsychological Assessment Devices (CNADs) for research and clinical purposes. The CNADs encompass any new computer-based neuropsychological assessments or computerised versions of already established paper-and-pencil tests (e.g., Wisconsin Card Sorting Test; Sahakian & Owen, 1992) or web-based tests. The CNAD could be a standalone device (i.e., hardware and software) or software (i.e., either installed locally or on the internet) that can be run on devices such as personal computers, laptops, tablets, or smartphones (Bauer et al., 2012). The VR-EAL, as an immersive VR software and neuropsychological assessment, would be categorised as a CNAD. Hence, VR-EAL or any other VR CNAD should meet the criteria of AACN and NAN to be effectively implemented for clinical or research purposes.

The AACN and NAN recognise the potential advantages of CNADs

**Scene 1**

**Scene 1**

**Scene 1**

**Scene 2**

**Scene 2**

**Scene 4**

**Scene 4**

**Scene 5**

**Fig. 1.** VR-EAL tutorials: Scenes 1-5.

which include testing large numbers of individuals quickly (e.g., parallel administration); immediately available tests; enhanced accuracy and precision (e.g., reaction time measurements); shorter administration time and reduced costs (e.g., for test administration and scoring); adaptable in different languages; exporting the data automatically (e.g., for research purposes); increased accessibility (e.g., remotely); and the integration of algorithms for making decisions on issues such as the identification of an impairment or a statistically reliable change (Bauer et al., 2012). In these series of studies, the VR-EAL has already shown that it achieves several of these benefits. The VR-EAL is immediately available after its installation on a personal computer and automatically produces accurate performance scores that are exported into a.txt file (Kourtesis et al., 2020b). Consequently, the VR-EAL has no costs for administration and scoring, and it requires a substantially shorter administration time as compared to the equivalent paper-and-pencil batteries (Kourtesis, Collina, et al., 2020).

However, the VR-EAL currently does not incorporate a predictive algorithm for identifying cognitive impairment, since it has not been administered to any clinical populations. Thus, the predictive validity of VR-EAL has yet to be established and this is one of our future directions. Furthermore, the procedure for adapting the VR-EAL for use with different languages and cultures is more complex than the adaptation of a paper-and-pencil test, since, in the case of the VR-EAL, this procedure requires programming and software development skills, which will necessitate more time. Lastly, the VR-EAL may be accessed remotely, yet the unsupervised (i.e., without a trained clinician or a researcher) administration of the VR-EAL is not recommended, and the installation requires hardware (i.e., immersive VR HMD, controllers, motion tracking devices, and a VR-ready personal computer) which may be unaffordable for an individual to purchase.

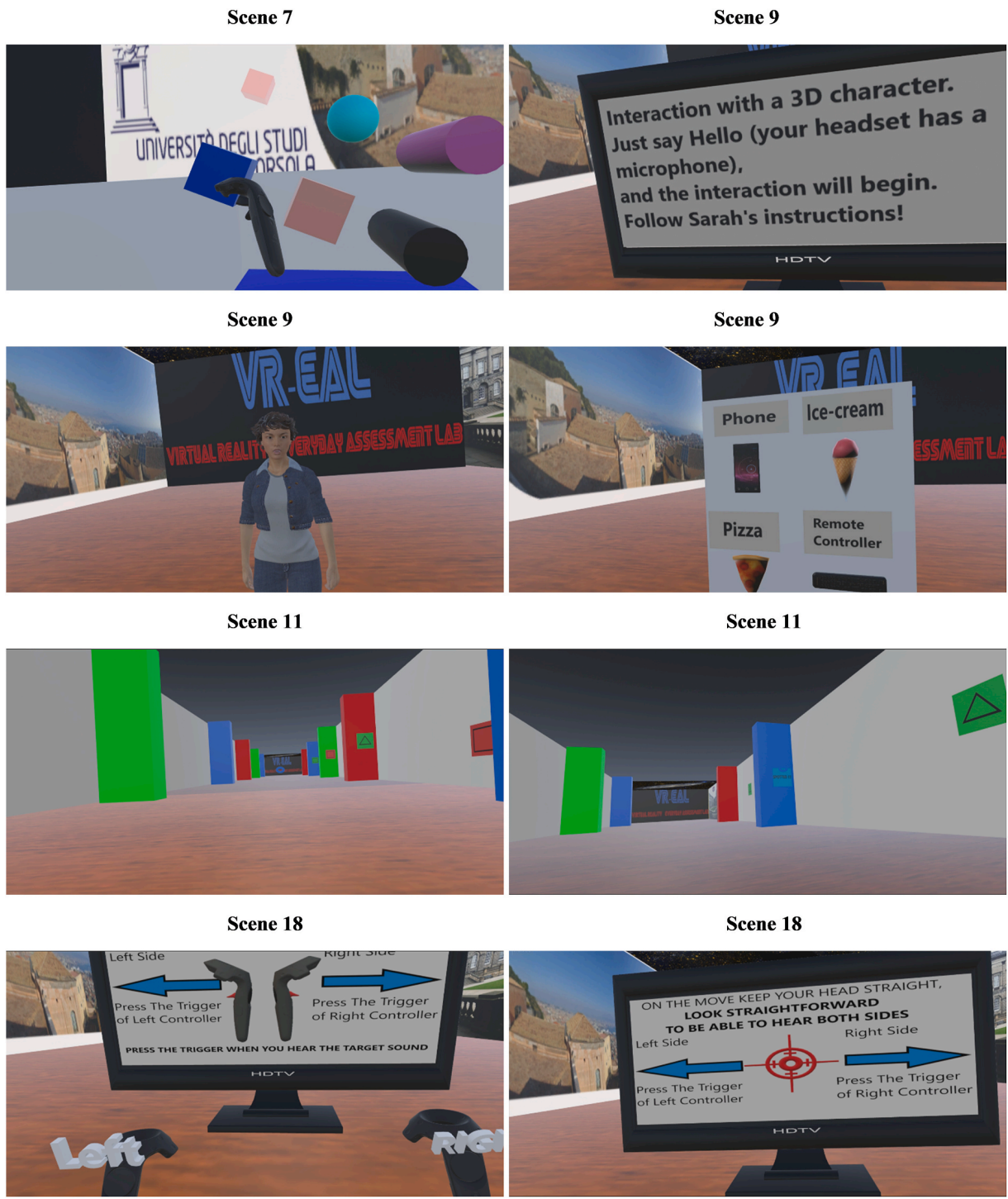

**Fig. 2.** VR-EAL tutorials: Scenes 7-18.

Nevertheless, as mentioned above, the AACN and NAN specified eight issues that should be addressed to benefit from the previous mentioned advantages of CNADs (Bauer et al., 2012). These issues are pertaining to: (1) the safety and effectivity of the CNAD; (2) the identity of the end-user (i.e., the operator of the CNAD); (3) the technical hardware and software features of the CNAD; (4) privacy and data security; (5) the psychometric properties of the CNAD; (6) examinee issues (e.g., cultural, experiential, and disability issues); (7) the use of reporting services; and (8) the reliability of the responses and results of the CNADs (i.e., the performance on CNADs; Bauer et al., 2012). Therefore, the utility of the VR-EAL should be discussed in relation to the guidelines for CNADs by AACN and NAN. The aim of this discussion is to highlight how the VR-EAL already satisfies these criteria, as well as to identify the shortcomings of the VR-EAL and define the necessary future

Scene 3 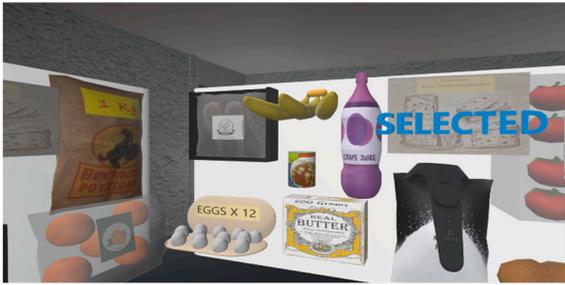 Scene 3 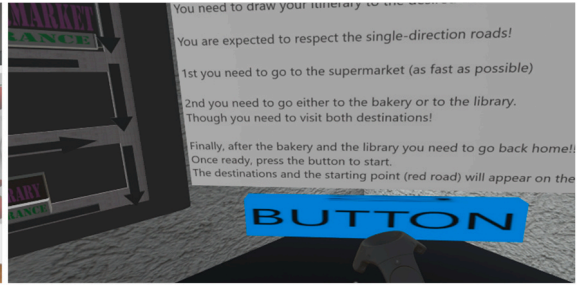

Scene 6 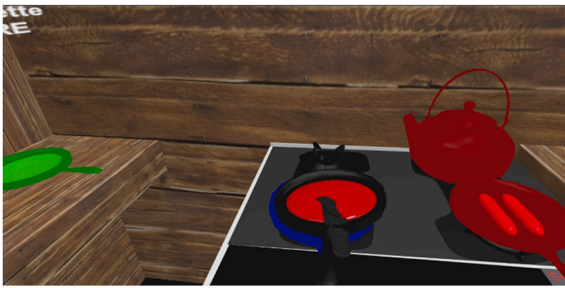 Scene 6 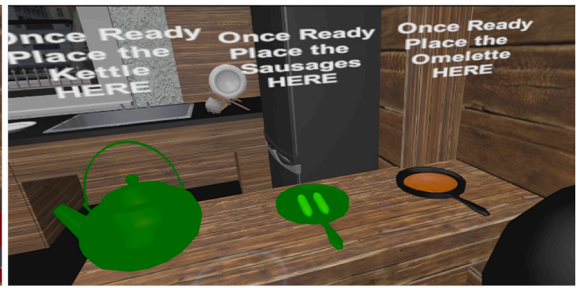

Scene 6 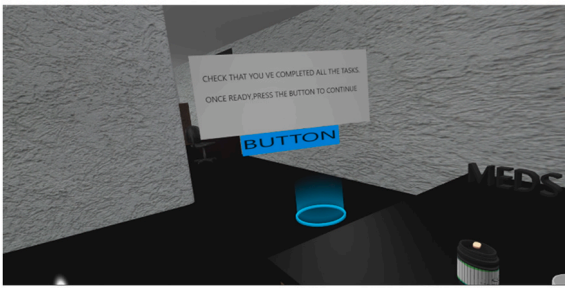 Scene 8 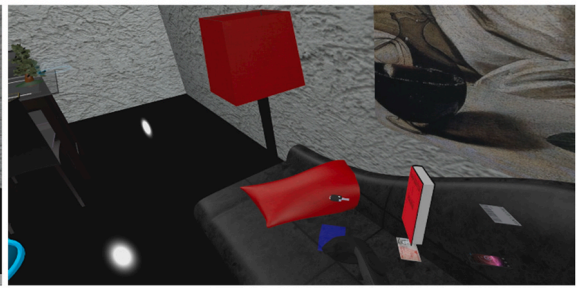

Scene 8 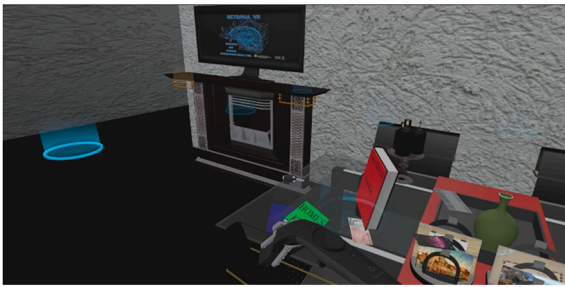 Scene 10 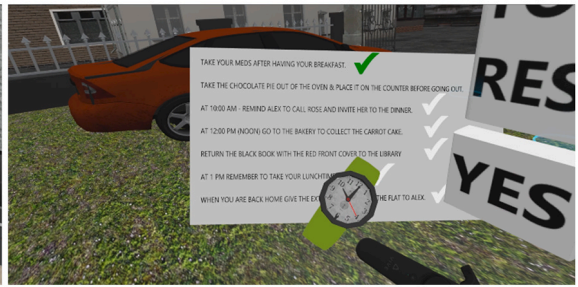

Scene 12 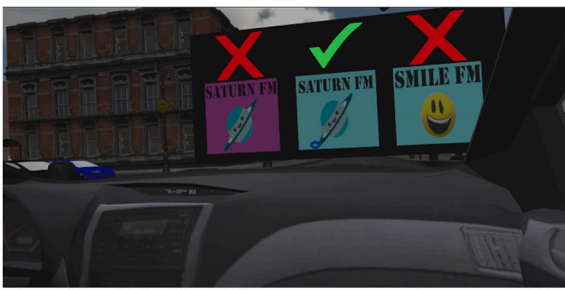 Scene 12 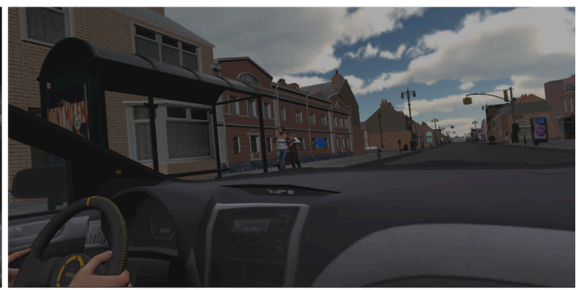

**Fig. 3.** VR-EAL storyline: Scenes 3-12.

**Scene 14** **Scene 14**

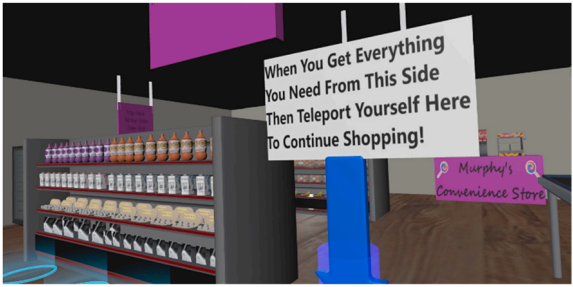 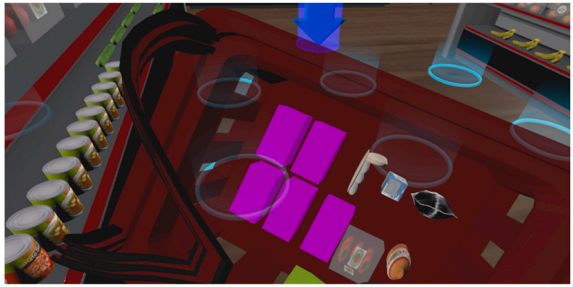

**Scene 15** **Scene 17**

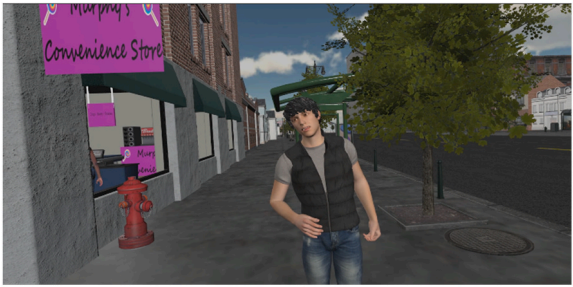 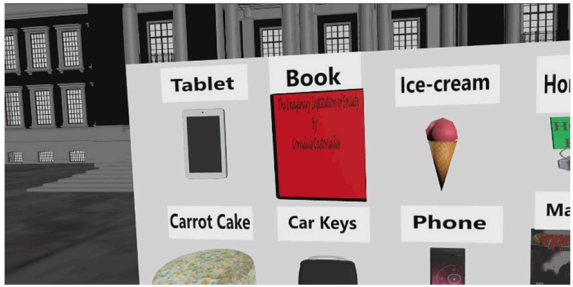

**Scene 19** **Scene 19**

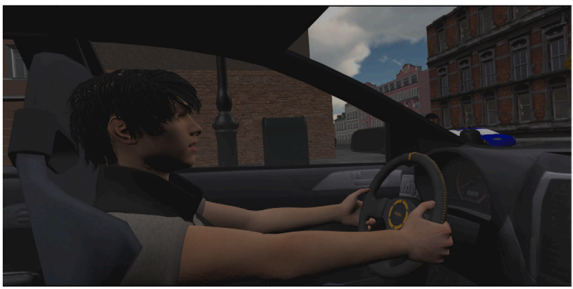 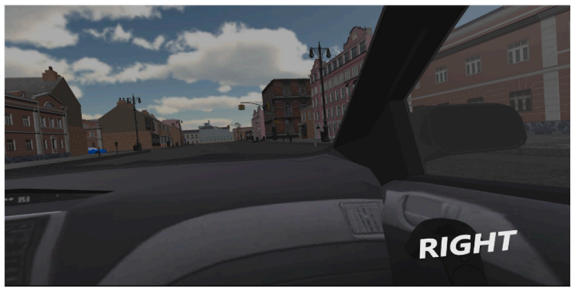

**Scene 20** **Scene 22**

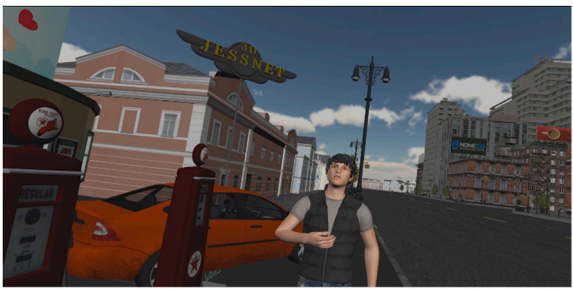 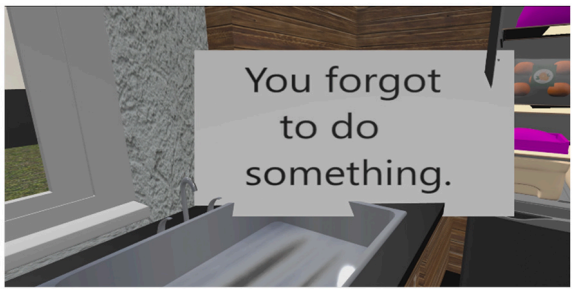

**Scene 22** **Scene 22**

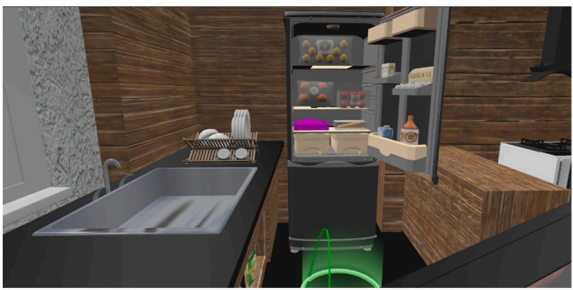 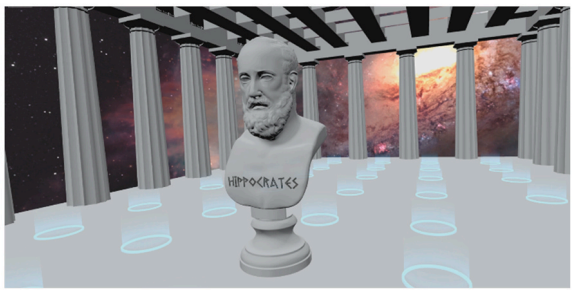

**Fig. 4.** VR-EAL storyline: Scenes 14-22.

**Table 2**
The main aims and findings of the studies.

| Aims | Findings |
|---|---|
| Identify the technical reasons for cybersickness and examine the effect of these technical factors in neuroscientific and neuropsychological studies. | • The review indicated features pertinent to display, sound, motion tracking, navigation mode, ergonomic interactions, user experience, and computer hardware that should be considered by researchers.<br>• The meta-analysis of the VR studies demonstrated that new generation HMDs induce significantly less cybersickness and marginally fewer dropouts. Importantly, the commercial versions of the new generation HMDs with ergonomic interactions had zero incidents of adverse symptomatology and dropouts. HMDs equivalent to or greater than the commercial versions of contemporary HMDs, accompanied with ergonomic interactions, are suitable for implementation in cognitive neuroscience. |
| Develop and validate the VRNQ and explore the maximum duration of VR sessions. | • VRNQ is a valid tool for assessing VR software in terms of self-reported user experience, game mechanics, in-game assistance, and cybersickness intensity; it has good convergent, discriminant, and construct validity.<br>• The maximum duration of VR sessions should be between 55 and 70 min when the VR software meets or exceeds the parsimonious cut-offs of the VRNQ, and the users are familiarized with the VR system. |
| Provide guidelines for the development of VR software in cognitive neuroscience and neuropsychology, by describing the development of VR-EAL. | • The Unity game engine, in conjunction with compatible software incorporating assets and software development kits, assist cognitive scientists in overcoming challenges pertinent to cybersickness and the quality of the VR software.<br>• Better in-game assistance, game mechanics, and graphics substantially increase the quality of the user experience and almost eradicate cybersickness.<br>• It is feasible to develop effective VR research and clinical software without the presence of cybersickness during a 60-min VR session. |
| Validate and compare VR-EAL against a paper-and-pencil ecologically valid neuropsychological battery. | • VR-EAL scores were significantly correlated with their equivalent scores on the paper-and-pencil tests.<br>• The participants' self-reports indicated that the VR-EAL tasks were considered significantly more ecologically valid and pleasant to perform than the paper-and-pencil neuropsychological battery. Also, the VR-EAL battery had a shorter administration time.<br>• The VR-EAL is a suitable neuropsychological assessment of everyday cognitive functions with enhanced ecological validity, providing a highly pleasant testing experience, and not inducing cybersickness. |
| Examine the focal and non-focal event-based, and time-based prospective memory using an ecological valid research paradigm, as well as to identify the cognitive functions which predict everyday prospective memory functioning. | • The length of the delay between encoding and retrieving the prospective memory intention, and not the type of prospective memory task, appears to play a central role in everyday prospective memory.<br>• Everyday prospective memory functioning is predominantly facilitated by episodic memory, visuospatial attention, and executive functions. |

directions for improving VR-EAL's utility as a research and clinical tool.

*3.1. End-user, privacy, and reliability issues (points 2, 4, 7, and 8)*

A critical issue is the targeted end-user of the CNAD (i.e., the person who operates the CNAD). As defined by the American Psychological Association (APA), researchers and clinicians "*do not promote the use of psychological assessment techniques by unqualified persons, except when such use is conducted for training purposes with appropriate supervision*" (APA, 2010, Ethical Standard 9.07, Assessment by Unqualified Persons). CNADs can be implemented by other professionals who do not have a background in psychometrics or neuropsychology but the results should be integrated and interpreted by a competent professional such as a cognitive neuroscientist or neuropsychologist (Bauer et al., 2012). Specifically, the VR-EAL should be administered by a clinician or researcher who has competency in both neuropsychological assessment and immersive VR technologies (Kourtesis et al., 2020b). Therefore, the definition that the end-user of VR-EAL should be a trained professional hence aligns with the ethical principles of the APA (APA, 2010, Ethical Standard 9.07, Assessment by Unqualified Persons).

Furthermore, regarding privacy and data security, test scoring and interpretation, and record keeping, the principal concern of AACN and NAN pertains to whether the end-user would be trained to follow the respective APA guidelines and ethical standards (Bauer et al., 2012). The cognitive neuropsychologist or neuroscientist administering the VR-EAL should abide with the record keeping guidelines (e.g., data should be stored and encrypted locally) of the APA (APA, 2007, Record Keeping Guidelines). The VR-EAL does not require an internet connection. Hence, the VR-EAL can be run offline, which eliminates the possibility of having the neuropsychological data accessed by possible digital perpetuators. The data that the VR-EAL provides should be thus handled and stored in the same way as other neuropsychological data that the clinician or researcher collects through traditional (e.g., paper-and-pencil test) methods. The VR-EAL offers a.txt file to the end-user, where all the recorded data (i.e., response times, duration of each task, quantification of various types of errors, and cognitive performance scores) are displayed (Kourtesis et al., 2020b). This.txt file and the containing data (e.g., if they have been transferred to an excel file) should be stored locally and encrypted, which is a common practice among researchers and clinicians (APA, 2007, Record Keeping Guidelines). Moreover, since the end-user of VR-EAL should be an individual trained in psychometrics and neuropsychology, the end-user should be capable of integrating and interpreting the data amassed by VR-EAL, which also agrees with the APA ethical standards for test scoring and interpretation (APA, 2010, Ethical Standard 9.09, Test Scoring and Interpretation Services). Therefore, the guideline that every VR-EAL end-user should be a cognitive neuropsychologist or neuroscientist meets points 2 (i.e., end-user issues), 4 (i.e., privacy and data security issues), and 7 (i.e., scoring and data recording issues) of the guidelines of AACN and NAN for the appropriate implementation of CNADs.

Furthermore, examinee cooperation and sufficient motivation are crucial for obtaining reliable neuropsychological test scores (AACN, 2007; Bauer et al., 2012; Heilbronner et al., 2009). Specifically, participants' efforts have been found to substantially affect performance on neuropsychological tests; indeed, in some studies, participants' effort was found to have a greater impact on their cognitive performance than the pathophysiological condition (Constantinou, Bauer, Ashendorf, Fisher, & McCaffrey, 2005; Stevens, Friedel, Mehen, & Merten, 2008; West, Curtis, Greve, & Bianchini, 2011). However, when the end-user of the CNAD is a trained clinician or researcher, they are capable of identifying behavioural signs (e.g., slow movements when there is not any motor disability) that there is reduced effort by the participant through behavioural observation (Bauer et al., 2012; Heilbronner et al., 2009). Nevertheless, the suspicion of poor effort on cognitive tests should be further explored and confirmed (e.g., using an effort test; Bauer et al., 2012; Heilbronner et al., 2009). Consequently, the suggestion that the end-user of the VR-EAL should be a trained clinician or researcher assists with the detection and confirmation of poor effort on the VR-EAL's tasks. In addition, the VR-EAL, as an immersive VR software which has game-like features (e.g., a user-centred interface) and simulates everyday tasks within a realistic scenario, appears to engage and motivate the examinees (Kourtesis et al., 2020a, 2020b). Notably, our two different samples of participants rated the VR-EAL as a highly pleasant testing experience (Kourtesis et al., 2020a, 2020b). Motivating the participant to perform the tasks is important for acquiring reliable data, while it also assists with identifying behavioural signs of poor effort (Heilbronner et al., 2009). Thus, the motivating nature of VR-EAL (i.e., a highly pleasant testing experience with an engaging scenario) may also assist with the avoidance or the detection of potential issues pertaining to the examinee's effort. Nevertheless, a future version of the VR-EAL should also consider hand and head movement speed to appraise whether the examinee is indeed motivated to engage with the VR-EAL. This system may assist end-users (clinicians and researchers) in detecting malingering or a lack of motivation.

### 3.2. Technical features, safety, and effectivity issues (points 1 and 3)

The AACN and NAN underline that a CNAD should meet the safety criteria of the Federal Food, Drug & Cosmetic Act (FD&C; Bauer et al., 2012). Section 201(h) of the FD&C (21 U.S.C. 301) defines a "*medical device*" as "*an instrument, apparatus, implement, machine, contrivance, implant, in vitro reagent, or other similar or related article, including a component part, or accessory which is … intended for use in the diagnosis of disease or other conditions, or in the cure, mitigation, treatment, or prevention of disease, in man or other animals . …*". Hence, a CNAD as a medical device should also comply with the safety criteria of FD&C (i.e., to not cause any harm to the examinees; Bauer et al., 2012). Any inconvenience or adverse effects may be attributed to the hardware and software features of the CNAD (Bauer et al., 2012; Cernich, Brennana, Barker, & Bleiberg, 2007). Likewise, the hardware and software features of a CNAD may compromise the effectivity of a CNAD and the reliability of the acquired neuropsychological and/or physiological data (e.g., Bauer et al., 2012; Cernich et al., 2007). Parsons, McMahan, and Kane (2018) argued that contemporary hardware (e.g., personal computers with dual processors) have the computing power to sustain the parallel operation of several software, while software are now developed to exploit and effectively use this computing power. These recent technological advancements pertaining to hardware and software, allow the parallel acquisition of accurate and reliable data such as reaction times, errors, neuroimaging data, and physiological data (Parsons, McMahan, & Kane, 2018). Regarding VR-EAL and VR CNADs, the principal problem is the presence of adverse cybersickness, which compromises the safety of the participants and the reliability of the acquired data. Individuals with conditions which affect the vestibular (e.g., stroke patients) and/or oculomotor systems (e.g., epileptic patients) may be susceptible to experience cybersickness or seizures respectively. Hence, the use of immersive VR should be avoided or excluded in individuals with vestibular and/or oculomotor symptoms or disorders. Regarding cybersickness, intense cybersickness has been found to compromise overall cognitive performance (i.e., neuropsychological data; Mittelstaedt, Wacker, & Stelling, 2019; Nalivaiko, Davis, Blackmore, Vakulin, & Nesbitt, 2015; Nesbitt, Davis, Blackmore, & Nalivaiko, 2017) and increase electrical activity and connectivity of frontotemporal and occipital lobes (i.e., neuroimaging data; Arafat, Ferdous, & Quarles, 2018; Gavgani et al., 2018; Toschi et al., 2017). The main cause of cybersickness is the implementation of immersive VR hardware (e.g., HMDs and personal computers) of inadequate quality (e.g., low resolution or processing power) and/or software that does not have certain features (e.g., ergonomic navigation and interaction system; Kourtesis et al., 2019a; 2019b).

A meta-analysis of VR neuropsychological and neuroscientific studies confirmed the importance of hardware characteristics for removing cybersickness, where studies that utilized contemporary HMDs had substantially less incidents of cybersickness and dropouts. Studies that used an HTC Vive HMD (Kim, Choe, Hwang, & Kwag, 2017) with two lighthouse stations for motion tracking (Plouzeau et al., 2015) and two HTC Vive wands with six degrees of freedom (6DoF) for navigation and interactions within the virtual environment (Figueiredo, Rodrigues, Teixeira, & Techrieb, 2018) have reduced or eradicated cybersickness. In line with this, the studies of Kourtesis and collaborators (2019b; 2020a; 2020b) used hardware that was in line with these hardware-related suggestions. There were no dropouts and the presence and intensity of cybersickness was minimal to none, which further confirms the importance of the suggested hardware characteristics. Labs and clinics can acquire an appropriate HMD since commercial desktop-based (e.g., HTC Vive) and standalone (e.g., Oculus Quest) HMDs can be purchased for a relatively low price (e.g., $300 - $500; Kourtesis et al., 2019a). As a result, recent immersive VR studies have implemented HMDs which meet the minimum hardware characteristics (e.g., Banakou, Kishore, & Slater, 2018; Detez et al., 2019; George, Demmler, & Hussmann, 2018; Mottelson & Hornnaek, 2017; Parsons & McMahan, 2017). Also, the VR-EAL is only compatible with these recent HMDs. Therefore, the VR-EAL appears to meet the hardware criteria of AACN and NAN, which ensure the safety of the examinees and the reliability of the acquired data.

Beyond the hardware characteristics, the quality of the software is also important to avoid or alleviate cybersickness incidence and intensity. Using an appropriate HMD and hardware when the software does not have the required characteristics may still result in intense cybersickness and dropouts (e.g., Detez et al., 2019). Navigation within the virtual environment should be facilitated by teleportation or physical movement or a combination of both (Porcino et al., 2017), and the interactions with the virtual environment should be ergonomic and naturalistic (Figueiredo et al., 2018). Furthermore, the in-game instructions, prompts, and tutorials should provide the user with adequate and salient information regarding the storyline, controls, and orientation (Jerald, LaViola, & Marks, 2017). Lastly, the audio and ambient sounds within the virtual environment should be spatialized and of high quality (Vorländer & Shinn-Cunningham, 2014). On the basis of these recommendations, the VR-EAL combines teleportation and physical movement as a navigation method (Porcino et al., 2017), provides haptic information and has ergonomic and naturalistic interactions (Figueiredo et al., 2018), spatialized and high definition audio (Vorländer & Shinn-Cunningham, 2014), and several informative in-game instructions, prompts, and tutorials (Jerald et al., 2017). In Kourtesis et al. (2019b), the cybersickness intensity was minimal and only related to fatigue. In Kourtesis et al. (2020a; 2020b), the implementation of VR-EAL showed no dropouts and the cybersickness incidence and intensity was negligible and again solely related to fatigue. These results replicated across all three studies confirming the significance of software features in avoiding or alleviating the incidence and intensity of cybersickness, as well as demonstrating that the VR-EAL incorporates these software features and does not induce significant cybersickness, which complies with the software criteria of AACN and NAN.

These VR software features are not only crucial for the avoidance or alleviation of cybersickness, but also for the efficiency of the VR software. The ultimate purpose of VR is to immerse individuals deeply enough to deceive the brain into believing that the virtual world is the real world. The depth of immersion depends on the strength of three perceptual illusions: the placement, plausibility, and embodiment illusions (Maister, Slater, Sanchez-Vives, & Tsakiris, 2015; Pan & Hamilton, 2018; Slater, 2009; Slater, Spanlang, & Corominas, 2010). The placement illusion is the deception that the virtual environment is a real one; hence, it depends on how close the virtual environment is to an equivalent real environment (Slater, 2009; Slater, 2018). The plausibility illusion is the deception that the virtual environment reacts to the laws of physics and the actions of the participant, thus, it depends on the proximity of the virtual environment's behaviour and senses to real life (Slater, 2009; Slater, 2018). The embodiment illusion is the deception that the virtual body of the participant is her/his own body; hence, it depends on the proximity of the virtual body's appearance and behaviour (i.e., synchronized with the movements in the physical environment) to the participant's real body and movement (Maister et al., 2015; Pan & Hamilton, 2018).

Beyond the level of immersion, the three illusions (i.e., placement, plausibility, and embodiment) are also important for the ecological validity of the immersive VR CNADs. The three illusions ensure that the individual will perform the tasks as s/he would perform them in real life (Maister et al., 2015; Pan & Hamilton, 2018; Slater, 2009; Slater, 2018). The VR-EAL has substantially strong placement and plausibility illusions, and a moderate embodiment illusion. Our participants reported that these illusions resulted in deep immersion levels (Kourtesis et al., 2020a, 2020b). Kourtesis and collaborators (2020b) provided an explicit description of how the VR-EAL tasks were designed to resemble everyday life tasks (e.g., cooking, shopping, and finding items in the living room). Notably, the participants reported that the VR-EAL tasks are very similar to the corresponding tasks that they perform in everyday life (Kourtesis et al., 2020a). Also, the performance of the participants on the VR-EAL tasks was significantly correlated with their performance on ecologically valid paper-and-pencil tasks. Hence, these two studies propose that the three illusions are crucial to ecological validity, as suggested by the previous literature (i.e., Maister et al., 2015; Pan & Hamilton, 2018; Slater, 2009; Slater, 2018). Thus, in line with the criteria of the Federal Food, Drug & Cosmetic Act, the software features of VR-EAL enabled the VR-EAL to efficiently achieve its purpose of delivering an ecological valid assessment of these everyday cognitive functions.

Finally, another safety measure is to provide an adequate space for immersion and naturalistic interaction within virtual environments (Borrego et al., 2018). In line with Borrego and collaborators' recommendation, the VR-EAL should be implemented in a VR area of 5 m$^2$, which should be clear of obstacles to allow the user to interact freely with the VR environment. Also, the VR-EAL embeds a chaperone (see SteamVR) which informs the user when participants approach the boundaries of the VR area. In addition, the VR-EAL has a teleportation system for navigation, which reduces the requirement for physical walking and the likelihood of falling on physical obstacles (e.g., a wall). Notably, the researcher or clinician should also be in the same room and supervise the examinee in order to intervene whenever appropriate (e.g., the user approaches the wall). Following these precautions, the VR-EAL facilitates a safe testing procedure.

### 3.3. Psychometric properties issues (point 5)

An important issue highlighted by the AACN and NAN is that, similar to traditional psychometric tests, the CNADs abide with the same standards and conventions of psychometric test development, such as providing evidence regarding their reliability, validity, and utility (Bauer et al., 2012). The information pertaining to the psychometric properties of the CNAD, which support the claimed purpose or application of the test, should be provided to potential end-users of the CNAD (Bauer et al., 2012). Notably, the APA ethical standards (APA, 2010) state that, "*Psychologists who develop tests and other assessment techniques use appropriate psychometric procedures and current scientific or professional knowledge for test design, standardization, validation, reduction or elimination of bias, and recommendations for use*" (Standard 9.05). Hence, all cognitive tests, either traditional or CNAD, must meet the minimum psychometric standards for reliability and validity. The validity of a test examines different psychometric properties of the test such as the content validity (i.e., the test measures the cognitive domain that is supposed to measure; e.g., episodic memory), construct validity (i.e., the test measures the cognitive function(s) that it is supposed to measure), and criterion-related validity (e.g., diagnostic validity, the test efficiently detects a cognitive disorder such as Alzheimer's disease; Nunnally & Bernstein, 1994). Similarly, the aspects that are examined for the reliability of a test are the internal consistency (i.e., the consistency across all the items of the test), rest-retest (i.e., consistency over time), alternate forms (i.e., consistency across all forms/versions of the test), and inter-rater reliability (i.e., consistency of the scores across diverse examiners; Nunnally & Bernstein, 1994). Importantly, as APA (2010) Ethical Standard 9.02 (Use of Assessments), Section (b) states, "*Psychologists use assessment instruments whose validity and reliability have been established for use with members of the population tested. When such validity or reliability has not been established, psychologists describe the strengths and limitations of test results and interpretation.*"

In VR-EAL, the principal aim was to develop an immersive VR neuropsychological battery with enhanced ecological validity for the assessment of cognitive functions central in everyday functioning. Hence, VR-EAL had to be consistent with the available ecologically valid assessments of these everyday cognitive functions. For the development of VR-EAL, the procedures and scoring systems of established ecologically valid paper-and-pencil tests such as the Test of Everyday Attention (Robertson, Ward, Ridgeway, and Nimmo-Smith, 1994), the Rivermead Behavioral Memory Test – III (Wilson, Cockburn, & Baddeley, 2008), the Behavioral Assessment of the Dysexecutive Syndrome (Wilson, Evans, Emslie, Alderman, & Burgess, 1996), and the Cambridge Prospective Memory Test (Wilson et al., 2005) were meticulously studied. However, the fact that the development of VR-EAL was based on the procedures and scoring of established ecological valid tests does not ensure that the VR-EAL will have equivalent psychometric properties. As the AACN and NAN suggest, even a computerised version of an established paper-and-pencil test should be treated as a new test, for which validity (e.g., content and construct validity) should be examined and confirmed (Bauer et al., 2012). For this reason, the psychometric properties of VR-EAL were assessed, where performance on the VR-EAL tasks significantly correlated with performance on the equivalent ecologically valid tests, which also supported the construct and content validity of the VR-EAL to assess these everyday cognitive functions (Kourtesis et al., 2020a). In addition, the VR-EAL tasks were rated by participants as more substantially more ecologically valid than the corresponding tasks of these tests, which may be attributed to the benefits of using immersive VR methods (Kourtesis et al., 2020a).

Overall, the content, construct, and ecological validity were explored and supported (Kourtesis et al., 2020a). Additionally, the VR-EAL has shown good internal consistency (i.e., reliability; Kourtesis et al., 2020a). Furthermore, high inter-rater reliability is warranted in every VR CNAD. Since the VR-EAL has a standardised and automated scoring method, there are not any differences across diverse end-users (i.e., the VR-EAL has perfect inter-rater reliability). However, since the VR-EAL does not have alternate forms, alternate-form reliability and test-retest consistency were not examined. The test-retest reliability of VR-EAL should be explored in future work. As both the validity and reliability of a test are not unitary psychometric properties, they should be re-examined as populations and the testing context changes over time (Nunnally & Bernstein, 1994). Notably, the eventual aim of CNADs, such as VR-EAL, is their utilisation for research and clinical purposes in healthy aging and clinical groups such as dementias (Anderson & Craik, 2017), attention-deficit/hyperactivity disorder and autism (Karalunas et al., 2018), mild cognitive impairment (Schmitter-Edgecombe et al., 2009), acquired and traumatic brain injuries (Groot, Wilson, Evans, & Watson, 2002), HIV (Woods et al., 2008), schizophrenia (Twamley et al., 2008), and Parkinson's disease (Pirogovsky et al., 2012). Thus, the administration of the VR-EAL in healthy aging and clinical populations may highlight its clinical utility through an exploration of its diagnostic validity (e.g., in the detection of mild cognitive impairment) and predictive validity (e.g., predicting everyday functionality and the independence of older adults).

One limitation of this series of studies is that the VR-EAL was only administered to healthy young adults (18–45 years old) who are unlikely to demonstrate any cognitive impairments or disorders (Chaytor & Schmitter-Edgecombe, 2003). Hence, the validity of the VR-EAL should also be studied in older adults and it may elucidate issues associated with cognitive ageing. For example, an age-related paradox is observed in relation to prospective memory and older adults, where older adults are impaired on laboratory-based prospective memory tasks, but perform better than younger adults on naturalistic tasks (Schnitzspahn, Ihle, Henry, Rendell, & Kliegel, 2011). Due to their increased life-experience and crystallised intelligence, older adults appear to be more effective in using environmental cues and compensatory strategies such as having a structured plan of action (e.g., noting down the sequence of necessary tasks), setting reminders (e.g., using notes, alarm clocks, or smartphones), making stronger and more complex associations between a task and an environmental cue (e.g., seeing a building, which used to be a post-office in the past, may remind them of the intention to mail a postcard to a relative), and using specialised items (e.g., using a dosette box to manage medications; Chaytor & Schmitter-Edgecombe, 2003; Marsh, Hicks, & Landau, 1998; Schnitzspahn et al., 2011). However, the utilisation of such techniques is not feasible in non-ecologically valid tests because their structured procedures only allow participants to respond or perform the task in a certain way (e.g., pressing a button on the keyboard, when seeing a specific item on the screen; Marsh et al., 1998; Schnitzspahn et al., 2011).

Consequently, the prospective memory age-related paradox highlights the importance of ecological validity in the assessment of everyday cognitive functioning (Chaytor & Schmitter-Edgecombe, 2003; Schnitzspahn et al., 2011). However, tasks performed in the real world (e.g., Marsh et al., 1998) cannot be standardized to allow their administration in other clinics or laboratories (Parsons, 2015). Also, they may not be appropriate for some individuals in challenging populations (e.g., a patient using a wheelchair), they are time-consuming and expensive (e.g., they require participant transport and consent from local businesses), and they do not have experimental control over the external situation (Parsons, 2015). In contrast, immersive VR CNADs like VR-EAL enable an adequate level of experimental control, while they are more cost-effective and inclusive than real-world tasks (i.e., naturalistic tasks; Parsons, 2015).

Potentially the VR-EAL could be used in the future to investigate the age-related paradox in prospective memory functioning in older and younger adults. This may also clarify the veridicality and predictive validity of VR-EAL by examining the existence of potential relationships between VR-EAL scores and established questionnaires assessing the ability to perform instrumental activities of daily life. Also, the inclusion of patients with mild cognitive impairment, which is a challenging population for diagnostic cognitive tests (i.e., tests frequently fail to achieve an adequately high sensitivity and specificity in differentiating individuals with mild cognitive impairment from healthy controls; Schmitter-Edgecombe et al., 2009), may inform on the predictive validity of VR-EAL by examining its sensitivity and specificity in differentiating older adults with mild cognitive impairment from healthy older adults. In summary, in line with the guidelines of AACN and NAN on providing evidence for a CNAD's utility (i.e., psychometric properties), this series of studies has demonstrated the ecological validity of VR-EAL, as well as its content and construct validity in young adults. However, the experimental and clinical utility of VR-EAL should be further explored in healthy older adults and individuals with major (e.g., dementias) or mild (e.g., mild cognitive impairment) neurocognitive disorders.

### 3.4. Examinee issues (point 6)

Another important concern of AACN and NAN regarding the implementation of CNAD is that individual differences (e.g., age, culture, education, motor abilities, and computer skills) may affect the examinees' performance on CNADs (Bauer et al., 2012). For these reasons, the developers of CNADs should investigate how diverse age, and cultural and educational backgrounds may affect the performance of the examinees, and then provide normative data correspondingly (Bauer et al., 2012). Furthermore, cognitive, motor, or sensory disabilities might have an impact on the examinees' ability to perform the CNAD's tasks effectively; hence, the suitability of the tests for individuals with disabilities should be explored and documented (Bauer et al., 2012). Finally, competency and familiarity with computers may also affect the validity of the CNAD's results (Bauer et al., 2012). Indeed, there are significant individual differences pertaining to the competency and familiarity with computer use (Iverson, Brooks, Ashton, Johnson, & Gualtieri, 2009). For example, gamers have been found to have faster perceptual processing speed compared to non-gamers, regardless their performance on tasks (e.g., number of errors and correct responses; Kowal, Toth, Exton, & Campbell, 2018). Importantly, the results from computerized versus paper-and-pencil tests may be substantially different in computer-familiarized versus computer-naive populations (Feldstein et al., 1999; Iverson et al., 2009).

However, the examinee's competency in using computers mainly influences performance on non-immersive CNADs. The user interface and procedure of non-immersive CNADs can be challenging for individuals without gaming backgrounds or familiarization with computers (Parsons et al., 2018; Zaidi, Duthie, Carr, & Maksoud, 2018), especially for older adults (Werner & Korczyn, 2012; Zygouris & Tsolaki, 2015). On the other hand, immersive VR CNADs appear to rely significantly less on gaming or computing ability than non-immersive CNADs (Bohil, Alicea, & Biocca, 2011; Parsons, 2015; Teo et al., 2016). The first-person perspective in conjunction with naturalistic interactions (i. e., close to real-life actions) assist non-gamers to perform comparable to gamers in immersive VR environments (Zaidi et al., 2018). Indeed, the findings of the series of Kourtesis and collaborators' studies indicated that the gaming ability of the examinee does not affect the utilisation of immersive VR technologies and performance on the VR-EAL. There was no significant difference between gamers and non-gamers in the duration of the VR session (Kourtesis et al., 2019b). Similarly, performance on the VR-EAL appeared to demonstrate no difference between gamers and non-gamers (Kourtesis et al., 2020a, 2020b). Finally, performance on the VR-EAL was not found to be affected by age or educational background. Therefore, the VR-EAL appears to be appropriate for the assessment of young individuals regardless of their educational background, age, or competency in using computers.

However, as discussed above, the VR-EAL should also be administered to older adults to investigate their attitudes towards VR-EAL, and whether their competency in computers affects their performance on the VR-EAL. Nevertheless, recent studies have found that older adults, after using immersive software, expressed a very positive attitude towards immersive VR technologies and rated immersive VR software as a highly pleasant experience, while they did not experience adverse cybersickness (Appel et al., 2020; Brown, 2019; De Vries, Van Dieën, Van Den Abeele, & Verschueren, 2018; Huygelier, Schraepen, van Ee, Vanden Abeele, & Gillebert, 2019). Also, the application of immersive VR software was feasible in older adults with lower-motor disabilities (Appel et al., 2020; Brown, 2019), as well as in older adults with various levels of cognitive impairments (i.e., mild, moderate, and severe; Appel et al., 2020). However, older adults were found to prefer and perform better on immersive VR software that has ergonomic and naturalistic interactions (De Vries et al., 2018). Furthermore, both younger and older adults showed an increased motivation to perform cognitive tasks in immersive VR rather than traditional paper-and-pencil tests (Corriveau Lecavalier, Ouellet, Boller, & Belleville, 2020). Finally, the performance of both younger and older adults on episodic memory tasks in an immersive VR CNAD were analogous to their performance on traditional paper-and-pencil episodic memory tests, indicating that the performance of both younger and older adults was not affected by their competency in using computers (Corriveau Lecavalier et al., 2020).

In this series of studies, the VR-EAL, which provides ergonomic and naturalistic interactions, was rated as a highly pleasant testing experience by younger adults, whose performance on the VR-EAL was substantially correlated with their performance on equivalent paper-and-pencil tests. Therefore, based on the findings of the aforementioned studies (i.e., Appel et al., 2020; Brown, 2019; Corriveau Lecavalier et al., 2020; De Vries et al., 2018), in conjunction with the findings of this series of studies, it may be hypothesized that the future implementation of VR-EAL in older adults with diverse functionality (i.e., healthy individuals, individuals with cognitive impairments and/or lower-motor disabilities) is feasible. Furthermore, the implementation of VR-EAL in older adults is expected to offer a pleasant testing experience without cybersickness and show equivalent psychometric properties regardless their gaming/computing ability. Nevertheless, future implementations of the VR-EAL in diverse populations will allow one to examine VR-EAL's psychometric properties, strengths, and limitations to create more detailed documentation to assist VR-EAL's end-users with implementing VR-EAL competently.

*3.5. Beyond the AACN & NAN criteria*

The NAN and AACN criteria were formulated concerning computerized neuropsychological assessment devices. However, at that time, the chances and challenges of VR or extended reality (XR) applications were not yet foreseeable. Thus, the guidelines and criteria should be updated to address the ethical and practical challenges specific to VR and XR applications. As discussed above, Kourtesis et al. (2019b) showed the maximum duration of VR sessions without the presence of cybersickness. In our later work, we provided guidelines for the development of VR software appropriate for neuropsychological research or clinical practices (Kourtesis et al., 2020b). Ethical and privacy implications in mixed reality applications (Bye, Hosfelt, Chase, Miesnieks, & Beck, 2019) and developing virtual worlds (e.g., VR applications), as well as implementing them in training, education, psychotherapy, rehabilitation, marketing, entertainment, and research (Slater et al., 2020) have also been discussed. Vasser and Aru (2020) provided guidelines for implementing immersive VR systems in psychological research. Finally, Krohn and collaborators (2020) developed a multidimensional evaluation framework for VR applications in clinical neuropsychology (a.k.a. VR-Check).

The VR-Check framework encompasses 10 principal evaluation dimensions, such as cognitive domain specificity, ecological relevance, technical feasibility, user feasibility, user motivation, task adaptability, performance quantification, immersive capacities, training feasibility, and predictable pitfalls. As Table 3 displays, there is indeed a comparability between the AACN and NAN criteria with the VR-Check framework dimensions. This could indicate that the VR-EAL also satisfies the dimensions-issues raised by Krohn et al. (2020). Furthermore, although the AACN & NAN Criteria can be used to evaluate a VR-CNAD like VR-EAL, VR-Check addresses issues pertaining specifically to VR software. A future joint position and guidelines for AACN & NAN may include the dimensions of the VR-Check Framework as well as the guidelines raised by other VR studies (e.g., Bye et al., 2019; Kourtesis et al., 2020b, 2019b; Slater et al., 2020; Vasser & Aru, 2020).

## 4. Limitations and future directions

This series of studies also have some limitations that should be considered. As the current series of studies aimed to explore the appropriateness and utility of the immersive VR methods in cognitive neuroscience and neuroscience, the various types of VR software (i.e., including VR-EAL) were only administered to younger healthy adults with a relatively high level of education. While performance on the VR-EAL was not found to be affected by age, education, or gaming ability, it would be important for these relationships to be examined in a more education- and age-diverse population including older adults. Also, technological means like immersive VR may further contribute in the diagnosis and care of major neurocognitive disorders like dementia (Moyle, 2019). As discussed above, the clinical and experimental utility of the VR-EAL should be further investigated in individuals with neurocognitive disorders, including Alzheimer's disease and mild cognitive impairment.

VR-EAL offers an assessment of prospective memory and relevant cognitive functions central to everyday functioning (Kourtesis, Collina, Doumas, & MacPherson, 2021; Kourtesis & MacPherson, 2021). Impaired prospective memory is frequently found in healthy older individuals (Huppert, Johnson, & Nickson, 2000), as well as in clinical populations such as Alzheimer's disease (Jones, Livner, & Bäckman, 2006), individuals with mild cognitive impairment (Costa, Caltagirone, & Carlesimo, 2011), and autism spectrum disorder (Sheppard, Bruineberg, Kretschmer-Trendowicz, & Altgassen, 2018). The ecologically valid examination of prospective memory may hence assist with improving our understanding of everyday prospective memory functioning, and consequently provide ways to improve functioning in individuals with impaired prospective memory. Also, administering the VR-EAL to both younger and older adults may help to probe the effect of aging on prospective memory. Older adults have reported a high acceptance of immersive VR systems (Huygelier et al., 2019; Syed-Abdul et al., 2019). These findings further postulate that the VR-EAL could be administered in older adults or individuals with neurocognitive disorders. Nevertheless, the clinical utility and predictive validity of VR-EAL has to be explored in future studies.

Furthermore, the VR-EAL presented some limitations as a CNAD in our series of studies. As discussed above, the VR-EAL induces strong placement and plausibility illusions, though, the embodiment illusion is only of moderate strength because it only relies on hand/controller movements. The embodiment illusion relates to the illusion of owning a virtual body (i.e., virtual avatar; Maister et al., 2015; Pan & Hamilton, 2018) and is important for acquiring cognitive and behavioural data, which resemble the individual's cognition and behaviour in real life (Maister et al., 2015; Pan & Hamilton, 2018). Despite this limitation, VR-EAL was rated as very similar to real life (i.e., enhanced ecological validity); however, improvement in the embodiment illusion would probably increase the already enhanced ecological validity of the VR-EAL. The most common and easy technique to create a responsive virtual body is the utilisation of software development kits (e.g., VRIK, Final IK, and IK for VR) which offer reliable and accurate inverse kinematics (i.e., animating the virtual avatar with respect to the user's movements; Lugrin et al., 2018).

Nonetheless, the virtual body should be as close as possible to the actual appearance and body of the examinee. Owning a virtual body that is dissimilar to the examinee's body may affect the performance of the examinee, either positively or negatively (Maister et al., 2015; Pan & Hamilton, 2018). For example, owning a virtual body which resembles that of Albert Einstein was found to significantly increase cognitive performance (Banakou et al., 2018). Regardless of whether the impact of the virtual body on cognitive performance is positive or negative, the virtual body that an immersive VR CNAD like VR-EAL offers should be as similar as possible to the examinee's body in order that the observed cognitive performance relates to the examinee's everyday cognitive ability. Hence, a future version of VR-EAL should include an application (e.g., using a photograph(s) of the examinee) which generates a virtual avatar that looks similar to the participant as in other immersive VR software (e.g., EngageVR).

**Table 3**
Comparability of AACN & NAN criteria with VR-Check framework.

| VR – Check Framework | AACN & NAN Criteria |
| --- | --- |
| Cognitive Domain Specificity | Psychometric properties (point 5) |
| Ecological Relevance | |
| Task Adaptability | Reliability Issues (points 7, 8) |
| Technical Feasibility | Technical Features (point 1) |
| User Feasibility | End-user issues (point 2) |
| User Motivation | Examinee Issues (point 6) |
| Performance Quantification | Reliability Issues (point 7) |
| | Privacy Issues (point 4) |
| Immersive Capacities | Technical Features (point 1) |
| Expected Pitfalls | Safety and Effectivity Issues (point 3) |
| Training Feasibility | NA |

Also, the VR-EAL does not have a mechanism to measure time monitoring, although there is a digital watch which the examinee uses to monitor time. Time monitoring has been found to be crucial in time-based prospective memory functioning (McFarland & Glisky, 2009; Mioni & Stablum, 2014; Vanneste, Baudouin, Bouazzaoui, & Taconnat, 2016). One way to implement time monitoring is to use the same gaze interaction system as in the VR-EAL visual attention task. Here, the gaze interaction system uses an invisible ray emitted (i.e., ray-casting) from the forehead point between the eyes (i.e., the upper point of the nose) straight towards the centre of the participant's field of view. This would allow for the recording of when and how many times the participant reads the time on the digital watch. The inclusion of a quantified time monitoring score in a future version of the VR-EAL will facilitate a more comprehensive assessment of time-based prospective memory.

Moreover, prospective memory components (e.g., retrospective), cue attributes (e.g., focality and salience of the cue) and the role of executive functions are important in prospective memory functioning, and they should be further explored in future studies. However, ecological valid CNADs like VR-EAL, which simulate everyday tasks (e.g., cooking), may be susceptible to confounding factors and fail to thoroughly examine a specific cognitive process. On the other hand, laboratory tasks are able to exclude confounding factors and permit the examination of a specific cognitive process. Nevertheless, laboratory tasks in their current form suffer from limitations such as their two-dimensional environment, non-naturalistic and non-ergonomic responses (i.e., using a keyboard, a button box, or joystick), static stimuli, and a substantial divergence from looking realistic. The utilisation of immersive VR technologies may be capable of resolving the limitations of laboratory tasks. Immersive VR laboratory experiments would facilitate a 360° testing environment, which incorporates realistic and dynamic stimuli, where the participant can interact in an ergonomic and naturalistic way (i.e., using wands/controllers or her/his own hands). Therefore, immersive VR laboratory experiments would minimize the divergence from real-life conditions, while facilitating a meticulous examination of the prospective memory components and cue attributes, as well as the role of executive functions in prospective memory functioning.

The VR-EAL can be utilized as an entire scenario for the assessment of everyday prospective memory, episodic memory, visual attention, visuospatial attention, auditory attention, and executive functions. However, the VR-EAL also offers a shorter scenario, where the aforementioned cognitive functions can be assessed without prospective memory and auditory attention. Moreover, the VR-EAL tasks may be administered independently (i.e., a generic tutorial, a specific tutorial for this task, and the storyline task) for the assessment of an individual cognitive function (e.g., multitasking, visual attention). Hence, VR-EAL could be considered as a mini library of immersive VR cognitive assessments. In the future, the sum of VR-EAL assessments (i.e., whole scenario, short scenario, and independent tasks) in conjunction with any future immersive VR CNAD may form an open access and source library of immersive VR software for cognitive neuroscience and neuropsychology. Considering the widespread adoption of open access and source tools such as PsychoPy (Peirce, 2007, 2009), OpenSesame (Mathôt, Schreij, & Theeuwes, 2012), R software (Culpepper & Aguinis, 2011), and the Psych Package (Revelle, 2011) in the last decade, the creation of an open access and source library of immersive VR software will promote the adoption of immersive VR technologies in cognitive neuroscience and neuropsychology.

Immersive VR technologies (i.e., HMDs) are compatible with electroencephalography (EEG; Teo et al., 2016), eye-tracking (Pettersson et al., 2018), and near-infrared spectroscopy (Teo et al., 2016). Neuroimaging and eye-tracking techniques been used in combination with immersive VR in several studies (e.g., Hofmann et al., 2018; Pettersson et al., 2018; Singh, Gramann, Chen, & Lin, 2021). Although combining neuroimaging and eye-tracking techniques with immersive VR has shown great potential for advancing psychological research, they have not been extensively implemented together (Pettersson et al., 2018; Teo et al., 2016). Future immersive VR software for implementation in cognitive neuroscience should strive to incorporate neuroscientific methods.

Eye-tracking may offer a detailed map with the trajectories of the examinee's eye gaze alongside response times (i.e., the time that the examinee's gaze fell on this point) and performance on the immersive VR CNAD (Pettersson et al., 2018). For example, combining eye-tracking with an immersive VR CNAD may assist with clarifying whether impaired performance on a cognitive task (e.g., abstract reasoning) is indeed due to an impaired ability on the assessed cognitive function or due to impaired attentional processes (Pettersson et al., 2018). Comparable to traditional approaches, combining neuroimaging techniques (e.g., EEG) with an immersive VR CNAD may inform on which brain regions are activated when a cognitive task is performed (Teo et al., 2016). Also, the combined implementation of immersive VR software with neuroimaging techniques such as EEG facilitates the utilisation of a brain computer interface (BCI; i.e., a direct communication pathway between the brain and an external device), where the examinee controls her/his virtual body in the virtual environment by activating predefined brain regions (Teo et al., 2016). For example, the examinee thinks the word "forward" to move her/his virtual body forward in the virtual environment. Using a BCI allows examinees with severe motor disabilities (e.g., tetraplegic) to perform tasks in an immersive VR CNAD (Teo et al., 2016). Hence, an open access and source library for immersive VR software in cognitive neuroscience should facilitate and/or incorporate some of the aforementioned neuroscientific methods (e.g., eye-tracking and EEG).

## 5. Conclusions

This series of studies endeavoured to address the shortcomings pertaining to the implementation of immersive VR technologies in cognitive neuroscience and neuropsychology by providing essential technological knowledge for the selection of appropriate hardware (i.e., HMDs, external, and computer) and software, as well as guidelines for the in-house and cost-effective development of immersive VR software. In addition, an advancement of the current available immersive VR research methods was attempted by developing and validating the VRNQ and VR-EAL. The VRNQ appears to be a valid and reliable tool for the appraisal of the intensity of cybersickness and the VR software features which are crucial for the alleviation or avoidance of cybersickness. The VR-EAL is the first immersive VR neuropsychological battery with enhanced ecological validity for the assessment of everyday cognitive functions, which facilitates a pleasant testing experience without inducing cybersickness. The VR-EAL was also found able to contribute to the understanding of everyday cognitive functions, which provides further evidence for the utility of immersive VR methods in cognitive neuroscience and neuropsychology. It is hoped that the findings of these series of studies have demonstrated the utility of immersive VR methods for improving the ecological validity and realism of neuropsychological assessment.

**Declaration of competing interest**

The authors declare no conflict of interest.

**Acknowledgments**

The VR-EAL is a free software that can be used by any clinician or researcher. PK is the developer of VR-EAL. Both authors were involved in the studies where VR-EAL was developed and implemented. The software review was conducted based on the objective criteria of the National Academy of Neuropsychology and the American Academy of Clinical Neuropsychology. The authors declare no conflicts of interest.